\documentclass[11pt]{article}
\usepackage{geometry}                % See geometry.pdf to learn the layout options. There are lots.
\usepackage{graphicx}
\usepackage{amssymb}
\usepackage{epstopdf}
\usepackage{verbatim}
\title{Optimizing the performance of Lattice Gauge Theory simulations with Streaming SIMD extensions}
\author{Shyam Srinivasan \\ shyams@uci.edu \\ Department of Information and Computer Science, \\ University of California, Irvine}
\date{}                                      % Activate to display a given date or no date

\begin{document}
\maketitle{}

\begin{abstract}
Two factors, which affect simulation quality are the amount of computing power and implementation. The Streaming SIMD (single instruction multiple data) extensions (SSE) present a technique for influencing both by exploiting the processor's parallel functionalism. In this paper, we show how SSE improves performance of lattice gauge theory simulations. We identified two significant trends through an analysis of data from various runs. The speed-ups  were higher for single precision than double precision floating point numbers. Notably, though the use of SSE significantly improved simulation time, it did not deliver the theoretical maximum. There are a number of reasons for this: architectural constraints imposed by the FSB speed, the spatial and temporal patterns of data retrieval, ratio of computational to non-computational instructions, and the need to interleave miscellaneous instructions with computational instructions. We present a model for analyzing the SSE performance, which could help factor in the bottlenecks or weaknesses in the implementation, the computing architecture, and the mapping of software to the computing substrate while evaluating the improvement in efficiency. The model or framework would be useful in evaluating the use of other computational frameworks, and in predicting the benefits that can be derived from future hardware or architectural improvements.
\end{abstract}
\section{Introduction}
\paragraph{}
Simulations allow scientists to observe phenomenon that are infeasible to recreate and inexpensively search parameter space. An important factor affecting simulation quality is computing power. More computing power - processing speed, and memory capacity - opens up the possibility of studying the problem in greater depth: either in terms of greater detail or bigger problem sizes. Parallel computing, which uses multiple components in concert to tackle larger computations is an especially effective way to increase computing power. The advent of parallelism at multiple levels of the computer hierarchy presents multiple avenues for optimizing scientific simulations. In this paper, we exploit the parallelism at the functional unit level of the processor through the use of the Streaming SIMD Extensions(SSE)~\cite{intel2} of the Intel Pentium 4 processor~\cite{intel} specifically for speeding up lattice gauge theory simulations~\cite{milc}. 
\paragraph{LGT, QCD and MILC for the layman}
The standard model of Physics consists of two quantum field theories: one which provides a unified theory of weak and electromagnetic interactions, and the other Quantum Chromodynamics (QCD), which provides a theory of strong interactions. Quantum Chromodynamics studies the nature of forces between quarks: the fundamental building blocks of matter. Gluons act as the carriers of the strong forces between quarks. QCD has been very successful in explaining a lot of data in high energy and nuclear experiments, and cosmic ray experiments. However, as it has been difficult to extract many of the predictions of QCD, the standard model is incomplete. One way of exploring QCD is through large scale numerical simulations using the frame work of lattice gauge theories~\cite{wilczek}.  Lattice gauge theory(LGT) simulations involve the formulation of gauge field theories on a space-time lattice~\cite{creutz}. The main purposes of LGT simulation are threefold: a quantitative understanding of physical phenomena, to determine the number of parameters of the standard model, and precision tests of the standard model~\cite{milc2}. One of the computational goals of the simulations is to reduce the finite lattice spacing parameter a, so that we can get continuum like numbers which more closely approximate nature. 
\paragraph{}
MILC or MIMD lattice Computation is one of the widely used suites for running Lattice QCD simulations. The MILC algorithms involve the use of a four dimensional space-time lattice to simulate QCD. The lattice is a four dimensional structure, with each site on a lattice being defined by a unique four dimensional tuple/coordinate system.
The MILC algorithm from a computing viewpoint can be split into a number of routines. The main computations in all these routines deal with the calculation and update of the link values at each lattice site based on the values of the neighbouring sites. The important point is that each site in the lattice has eight neigbhours: two in each one of the four dimensions. These links between sites are represented as (Special Unitary) SU(3) matrices. The interactions between sites involves the invocation of the various matrix routines outlined in Table~\ref{milc_ins}. It is easy to observe that the computational running time of Lattice simulations grows very fast. An important factor in using clusters is to keep the Price/performance ratio very small. There are two important dimensions to this endeavour: one is that the communication between components be very fast with low latency, the other is that the computational processing should be very fast and efficient. Our study's aim is to improve the latter, and by consequence improve overall computational capacity. 
\paragraph{}
The Streaming SIMD extensions and Streaming SIMD extension 2 instructions of the Intel Pentium 4 processor provide the user with the flexibility to execute arithmetic instructions that operate on multiple data units in parallel. SSE offers the opportunity to perform the four single floating point instructions in one cycle, while SSE2 offers the opportunity to perform two double precision instructions in one cycle. Our lattice simulations and SSE2 implementation are based on the MILC codes and the existing MILC SSE implementation.
\paragraph{}
We have modified our architecture, to exploit this new feature of the Intel processor and we have observed as much as a 1.4 times best case increase in performance.  The speed up however is below the expected speedup of 2. A critical factor influencing performance is the nature of memory accesses and the memory layout of the data structures. This has implications for the kind of software architectures that can optimally use the SSE features. We analyze the reasons behind the performance and present possibilities to further exploit such architectures. To better understand the nature of this performance improvement and other possible optimizations, we present an overview of the architecture. 

\section{Intel Architecture}
\paragraph{}
The Intel Xeon processor is based on the Intel Netburst architecture~\cite{intel}, which differs from the earlier Intel Pentium architectures significantly. Several improvements over earlier architectures have improved the pipeline, and favour advanced multimedia and computing intensive applications in science and engineering. The Streaming SIMD extensions 1 and 2 also called SSE1 and SSE2, are one such addition.
\paragraph{}
This section includes a brief overview of the components of the Pentium 4  Microarchitecture and their impact on system performance. The Pentium 4 or the Netburst architecture  comprises a computational unit, memory systems to store instructions and data, and a bus system to carry data between different memory hierarchy levels and the micro processing unit.  All these components have a direct impact on performance.
\subsection{Computational Units}
\paragraph{}
The computational units form the core of the Central Processing unit. The theoretical maximum computational capacity is determined by the speed of the microarchitecture arithmetic and logic computational units that comprise the central processing unit. There are two factors, which determine the computational throughput of the CPU. The number of computational units and their functionality, and the frequency of the CPU clock. The Pentium 4 has a floating point execution unit, a floating point move unit, two double speed ALU units, one integer unit, one load, and one store port. The clock speed of the Xeon execution units, is twice the frequency of the processor clock frequency. Thus, the Pentium 4's can execute two instructions per clock cycle. The SSE and SSE2 take advantage of this property to execute one 128 bit word operation every clock cycle. Thus,  a 2.4 GHz  processor can complete 4.8 Giga floating point operations of double precision operands every second.
\paragraph{}
Another factor influencing execution speed is the pipeline depth of the processor. The longer the pipeline , the more the possibilities of parallelizing instruction execution and thus, attaining a speed up. One of the enhancements of the Pentium 4 processor over its Pentium III predecessor is the longer instruction pipeline length that improves speed. On the flip side this also makes the architecture susceptible to performance hits due to branch mispredictions. A branch misprediction occurs whenever the processor encounters an instruction that involves transferring the program control to non contiguous memory location. Normally, the program counter in the absence of branch instructions moves sequentially executing instructions which are located in contiguous memory locations.  Additionally, a jump instruction or a procedure call would change the program counter variable to a non contiguous location, causing all the instructions loaded in the pipeline to be flushed. Thus, longer pipelines also have the drawback of being susceptible to branch mispredictions. This is countered in the Netburst architecture, by using better branch prediction schemes. 
\subsection {Memory Subsystem}
\subsubsection{Memory} 
\paragraph{}
There are several levels of memory in a computer system. The memory at the top of the hierarchy is closest to the processor, and also the fastest in terms of access speeds. As a result, the retrieval times for these data items from the processor is the shortest. These memories are smaller, as faster memories cost more due to the increased complexity of circuitry that delivers shorter retrieval times. Correspondingly, memory at the bottom of hierarchy is cheaper, thus offering more memory while at the same time also incurring the penalty of very high retrieval access times.  One of the main issues facing today's microarchitectures is making the memory access times match the computational speed of the processor. We need to have enough instructions to keep the execution units busy for every cycle. Increasing memory speeds and memory sizes at the top of memory hierarchy, reduces the processor-memory speed gap and makes it possible to approach ideal computational throughputs. The less than perfect observed computational throughputs, to a large extent are caused by the gap in memory fetch to processor execution speeds.
\paragraph{}
\subsubsection{Data and Instruction Caches}
\paragraph{}
The memory subsystem comprises the L1 cache, L2 cache, and main memory(RAM). The Intel caches are implemented as set associative caches. An associative cache, which is subdivided into cache lines, will hold data in the cache line corresponding to the data's line location in paged memory. The P4 architecture has two kinds of caches, which are classified according to their functionality: instruction and data. 
\paragraph{L1 cache} 
This is the cache, which is the located closest to the computational units, at the top of the memory hierarchy. It is a four way set associative cache with 64-byte cache lines. The cache size is 8k.
\paragraph{L2 cache} 
it is an eight way set associative cache of size 512k that can hold instructions as well as data. The cache line is 128 bytes in size and consists of two 64-byte sectors.
\paragraph{Instruction Trace caches}
The Trace cache is the instruction equivalent of the L1 data cache, for holding Instructions, and has a capacity of 12k $u$ops. All the instructions in the trace cache are stored in the form of basic micro-operations or $u$ops.
\subsection{The Bus system}
\paragraph{}
The bus system has a direct impact on performance as it is the communication channel for shuttling data and instructions between various memory hierarchies and the processor. As the speed or the width of the bus increases,  more data can be delivered in the same amount of time. The CPU is  kept occupied for more number of clock cycles. This process improves efficiency and results in execution time being reduced. 
\paragraph{}
To illustrate with an example, the Netburst architecture currently provides a 533 MHz Quad pumped Bus interface unit between L2 cache and main memory.  This bus frequency translates to 4 GB/s. On the other hand, the communication channel between L2 cache and L1 cache operates at 48 GB/s for a CPU frequency of 1.5. Since the L1 cache is connected to the processor the  latency for accessing data from the L1 cache is lower. The L1 cache has an access latency of 2 processor cycles for an integer load and 6 processor cycles for floating point/SSE loads. The L2 cache access latency is higher at around 12 processor cycles. In the case of an L2 cache miss, it takes 12 processor cycles to get to the bus and back within the processor. Additionally, a memory system access takes around 6-12 bus cycles, with each bus cycle being several multiples of the processor cycle depending on the architecture. For example, for a system with a 400 Mhz bus and a 2.4 GHz processor the ratio would be 6. 
\paragraph{}
It is evident that while the L1 cache accesses are the fastest, main memory accesses are the slowest. A huge penalty is paid for memory load instructions which require data retrieval from memory. Such instructions result in the CPU being idle for several clock cycles leading to inefficient programs. To counter delays due to accesses that require data from main memory, the architecture has a hardware prefetcher at the L2 cache level, which tries to stay 256 bytes ahead of the current data location. One of the solutions from a programming viewpoint is to make sure that memory access patterns are as regular as possible.   
\subsection{Miscellaneous Microarchitecture Units}
\paragraph{}
Besides the basic components mentioned above, there are a few other units which help the smooth functioning of the system. 
\paragraph{Branch Predictor}
The P4 architecture has a much deep pipeline that increases overall speed. However, a deeper pipeline also causes greater penalties due to branch mispredictions. To counter this occurrence, the Netburst architecture makes use of a branch prediction unit. The P4 has a branch prediction unit BTB that is connected to the Trace cache, and a Front end BTB  that is connected to the L2 instruction cache through an Instruction prefetcher. BTB stands for branch target buffer, which holds the history information of all branches~\cite{x86}.
\paragraph{Out of order Execution}
The out of order execution engine is used for executing independent instructions in parallel. The existence of multiple functional units presents the possibility of executing several instructions simultaneously, as long as there are no data dependencies between the instructions. Such parallel instruction execution is based on the concept of speculative execution, where instructions may not be executed in the serial order in which they are encountered in the program. With speculation, we speculate on the outcome of branches, executing the program as if our guesses were correct~\cite{henpat}. For instance, we may have two instructions A and B that share no data dependencies, with A occurring before B in a sequential execution of the program. As there are no dependencies, instruction B can be executed by the processor irrespective of A's execution status. 
\paragraph{}
The out of order execution engine implements the task of out of order execution. The engine speculatively schedules instructions around delayed instructions, as long as the instructions do not share data dependencies. The engine is complemented by the Retirement logic unit which ensures that instructions are retired or committed in proper program order. The Out of order unit also improves the working of the Branch prediction unit by training the BTB on the latest branch history information. Thus the Out of order engine makes use of the concept of speculative execution to exploit the processor functional unit parallelism and improve performance.
\paragraph{Prefetcher} 
Due to the gap between processor and wire speeds, there is a substantial performance hit whenever the processor has to wait for instructions or data to arrive from memory. To offset this problem, the netburst architecture stays ahead of data accesses by prefetching 256 bytes of data into the L2 cache. 
\paragraph{Instruction Decoder}
As the IA-32 instructions are computationally expensive to decode, an instruction decoder converts these to basic operations called ($u$ops) micro-operations. The program is stored in the form of these $u$ops in the execution trace cache. The execution core executes the uops supplied to it by the Trace cache.
\subsection{Software design inferences}
\paragraph{Alignment}
The L1 cache is organized as consecutive blocks of 64-bits, called lines. This matches the arrangement of main memory, which is also arranged as consecutive blocks of 64 bits. Unaligned data refers to data, which straddle two consecutive 64-bit lines. Whenever a memory load operation is issued, the 64-bit line which contains the data, is loaded into the cache. Furthermore, memory loads always retrieve 64-bits of data starting from the requested memory location, in one fetch cycle. Unaligned data reduce computational performance as they could require two memory loads rather than one, if they straddle a 64-bit boundary. The problem can be combated by the process of alignment. Alignment is done by making all data structures start at memory locations, which are multiples of 8 bytes. Alignment of data is thus an important performance optimization to keep in mind while programming for speed.
\paragraph{Hints for design}
Users of the Netburst architecture, should optimize their algorithms and implementations to reduce the FSB bottleneck between processor and main memory, increase regular access patterns which make use of the Netburst's prefetch architecture and use features like SSE/SSe2 that allow multiple instructions in one clock cycle.
\section{Streaming SIMD extensions}
\paragraph{}
Streaming SIMD extension also known as SSE is the latest feature offered by Intel to users of high performance applications. Currently SSE comes in two flavours for 32-bit architectures, SSE and SSE2. Both flavours allow the application to perform arithmetic, data movement and other instructions on 128 bits of data in one clock unit or cycle. For e.g, a normal 32-bit add operation, which is one of the most basic operations possible, takes one clock unit. The 32-bit number could be an integer or a single precision floating point number. SSE now lets us do a 128 bit add operation in one clock unit. Thus, it is possible for us to do four single precision floating point or four integer or two double precision operations in one clock unit.
\paragraph{}
There are eight SSE registers xmm0 to xmm7, and all operations are performed by moving the required data from memory to the SSE register set. The access time for these registers is the same as the access time for normal registers and MMX registers. The various instructions possible with SSE can be categorized as data movement, arithmetic, logical, shuffle, unpack, cache control, status, and mathematical functions. All normal operations have SSE analogues. Each one of the xmm registers is a 128-bit register as opposed to normal registers, which are 32-bit registers and MMX registers, which are 80-bit registers. The 128-bit word is composed of a lower 64-bit word and a higher 64-bit word. The low word addresses the first 64 bits and the high word addresses bits 65 to 128. SSE instructions can be classified according to the way they use the high and low words of the xmm registers, as either packed or scalar operations. Packed instructions operate on both high and low words, while scalar instructions operate on the low word leaving the high word unchanged.
\paragraph{}
The latest features of the SSE suite also allow us to access SSE functionality through the use of intrinsic functions, vector classes, and compiler options. 
\section{Nature of the Algorithm}
\paragraph{}
The MIMD lattice programs~\cite{milc} comprise a variety of lattice applications. The MILC applications have a high percentage of double and single precision floating point computations. The SSE and SSE2 features of the Intel processor present a good opportunity to perceptibly increase the speed and efficiency of these applications. The programming techniques for SSE and SSE2 are similar. There is a near one to one correspondence between SSE and SSE2 instructions. The differences spring from the fact that the 128 bit manipulations in the SSE case operate on 4 single precision operands and in the SSE2 case on 2 double precision operands. Both SSE and SSE2 MILC routines have the similar algorithms for similar operations. The term SSE will be used for SSE as well as SSE2 unless otherwise stated. 
\paragraph{} 
The Lattice QCD programs computational operations comprise a set of functions which involve various operations on 3x3 matrices and 1x3 vectors. Though, they form a small portion of the code, their execution time as a percentage of program time is very high, close to the 80-90 \% range. So, direct optimization of these routines through SSE will improve execution time.

The MILC application SSE routines comprise 15 computational routines, which are invoked most often. The table below lists the computational routines along with a short description.

\begin{table}[htdp]
\caption{Description of SSE and SSE2 MILC routines}
\begin{center}
\begin{tabular}{|| l | p{11cm} ||}	
	\hline
	{\bf Function}	&	{\bf Description}	\\	\hline
	add\_su3\_vector	&	Adds two su3 vectors A and B	\\	\hline
	mult\_adj\_su3\_mat\_hwvec	&	Multiply a Wilson half-vector A by the adjoint of a matrix B	\\	 \hline
	mult\_adj\_su3\_mat\_vec	&	Multlpies an adjoint matrix A by a vector B	\\	\hline
	mult\_adj\_su3\_mat\_vec\_4dir	&	Multiply an su3\_vector A by an array of four adjoint su3\_matrices B,and store the result in an array of four su3\_vectors		\\	\hline
	mult\_adj\_su3\_mat\_4vec	&	Multiply the adjoint of each of four input matrices by an input vector, storing the resulting vectors in 4 separate vectors	\\	\hline
	mult\_su3\_an	&	Multlpies an adjoint matrix A by a matrix B	\\	\hline
	mult\_su3\_mat\_hwvec	&	Multiply a Wilson half-vector A by a matrix B	\\	\hline
	mult\_su3\_na	&	Multlpies a matrix A by an adjoint matrix B	\\	\hline
	mult\_su3\_nn	&	Multlpies a matrix A by a matrix B	\\	\hline
	mult\_su3\_mat\_vec		&	Multlpies a matrix A by a vector B	\\	\hline
	mult\_su3\_mat\_vec\_sum\_4dir	&	Multiply the elements of an array of four su3\_matrices by the four su3\_vectors, and adds the results to produce a single su3\_vector	\\	\hline
	scalar\_mult\_add\_su3\_matrix	&	Muliplies a matrix B by a scalar s and adds the result to a matrix A	\\	\hline
	scalar\_mult\_add\_su3\_vector	&	Multiplies a vector B by a scalar S and adds the result to a vector A	\\	\hline
	su3\_projector	&	It is the outer product of A and B, C$_{ij}$ = A$_i$ * B\_adjoint\_j	\\	\hline
	sub\_four\_su3\_vecs	&	Subtracts four vectors from vector A	\\	
	\hline	 
\end{tabular}

\end{center}
\label{milc_ins}
\end{table}
\paragraph{}
The most basic and common operation involves the multiplication of a complex vector by another complex vector. Each element of the vector, B$_{j}$ and the matrix, A$_{ij}$ represents the value at row $i$ and column $j$. The element is a complex number composed of a real and an imaginary part. Both real and imaginary parts could either be single or double precision floating point numbers. We illustrate the use of the SSE functionality with a matrix-vector multiplication. Other calculations can be viewed as more complex mathematical constructs of the basic vector-vector multiplication. 
\paragraph{}
The matrix-vector multiplication routine could be subdivided into 3 computational sections for analysis. Calculating the C$_{ij}$ ( C is the vector in which the result will be stored ) element forms one computational section. We compute the real and imaginary parts for each C$_{ij}$ element and store it in memory. The code below shows the assembly instructions for calculating the first element of the result vector of an adjoint matrix-vector multiplication. The example below illustrates the use of SSE2 for LGT codes. This example can be extended to SSE with very minor modifications to the program structure. The SSE routine would have similar operations with suitable instruction substitutions for handling single precision numbers. 
\paragraph{}
The adjoint matrix A is a 3x3 matrix and the vector B is a 3x1 vector. Each element is a complex number which occupies 128-bits i.e. one xmm$_n$ or SSE2 register. Each matrix row and vector column has 3 elements and take up three SSE2 registers each. The remaining 2 of the 8 SSE2 registers are used as a scratch pad for storing intermediate results.
\verbatiminput{assemb_prog}
\paragraph{}
The assembly segment above calculates the first element C$_0$ of the Vector C. For ease of analysis, the assembly segment is broken into 6 sections. The segment above computes the calculation shown below.
\begin{tabbing}
C$_0$.real \= = \= A$_{00}$.real x B$_0$.real + A$_{00}$.imag x B$_0$.imag +	\=      \ \ \ \ \ \ $ \leftarrow$ (1.1)\\
		   \>    \> A$_{10}$.real x B$_1$.real + A$_{10}$.imag x B$_1$.imag + 	\>      \ \ \ \ \ \ $ \leftarrow$ (1.2)\\
		   \>    \> A$_{20}$.real x B$_2$.real + A$_{20}$.imag x B$_2$.imag 	\>      \ \ \ \ \ \ $ \leftarrow$ (1.3)\\		
C$_0$.imag \= = \= A$_{00}$.real x B$_0$.imag - A$_{00}$.imag x B$_0$.real +	\= \ \ \ \ \ \ $ \leftarrow$ (2.1)\\
		     \>    \> A$_{10}$.real x B$_1$.imag - A$_{10}$.imag x B$_1$.real +	\> \ \ \ \ \ \ $ \leftarrow$ (2.2)\\
		     \>    \> A$_{20}$.real x B$_2$.imag - A$_{20}$.imag x B$_2$.real	\> \ \ \ \ \ \ $ \leftarrow$ (2.3)\\ 		 			      
\end{tabbing}

\paragraph{}
When a function call is made, the compiler inserts instructions for saving the current state of the stack and registers. Block 1 saves the status of the stack and registers, before starting the routine. This block is complemented by a block at the end of the routine, which reverses the actions of block 1 and restores the stack and registers.
\paragraph{}
Block2 move instructions (mov and movupd), handle the transfer of the first row of Matrix A and Vector B, into the SSE registers. From the calculation above, it is clear that each element of the Matrix A row and the Vector B, is used twice. Consequently, it makes sense to retain the elements of either the Matrix row or the Vector after they have been used for the first time. It is not possible to retain both the row and the vector column as there are only 8 SSE registers. It is better to retain the vector elements as they will be needed for the entire routine, as opposed to the matrix row elements which are only needed for a particular C$_i$.
\paragraph{}
The calculation of each C$_i$ involves 12 multiplication and 10 addition operations. Each complex and real part individually involves 6 multiplications. Block 3 loads the real part of A$_0$ into the high and low words of the xmm$_1$ SSE register, and the imaginary part of A$_0$ into the high and low words of the xmm$_2$ register. Block 4 carries out the operations to compute the calculation shown in lines (1.1) and (2.1) above. The (1.1) and (2.1) line computations are carried out by multiplying the contents of  xmm$_1$ (A$_{00}$.real,A$_{00}$.real) with xmm$_3$ (B$_0$.imag,B$_0$.real), and the contents of xmm$_2$ (A$_{00}$.real,A$_{00}$.real) with xmm$_3$ (B$_0$.imag,B$_0$.real). The results are stored in xmm$_1$ and xmm$_2$, keeping the contents of xmm$_3$ intact.
\paragraph{}
Similarly, blocks 4 and 5 compute lines (1.2), (1.3), (2.2), and (2.3) of the calculations shown above. We also add all the multiplication results and store the results in the xmm$_1$ and xmm$_2$ registers. Finally in block 6, we have half the contents of the real and imaginary parts in xmm$_1$ and xmm$_2$. The xmm$_1$ register's lower word contains the real half while the upper word contains the positive imaginary half. The xmm$_2$ register contains the negative imaginary half in the lower word and the real half in the upper word. So, in block 6, we swap the lower and upper words of xmm$_2$, negate the upper word, which is the imaginary half and add it to xmm$_1$, to obtain C$_0$, the first complex number of the C vector. 
\paragraph{}
The example above represents exactly a third of the computation needed for the multiplying the adjoint of a matrix with a vector. As the result involves the calculation of a 3 x 1vector, the overall computation comprises two more sections exactly similar in computational structure and efficiency, to the section above. A matrix-vector multiplication involves 36 multiplications and 30 additions.  Besides computational operations it also involves 27 moves, 15 shuffle , and 6 subtraction operations. As SSE2 can manage two arithmetic instructions in the time taken for one, the time taken to multiply, add and subtract is reduced by half to 18,15 and 3. The time taken to store and load however remain the same as we still have to move the same amount of data between processor and memory.   
\paragraph{}
The basic computational operation explained above can be extrapolated for other routines such as matrix-matrix calculations. Besides adjoint matrix-vector calculations there are several other routines, which involve matrix-matrix, matrix-vector, vector-vector, and vector/matrix-scalar operations. The example above was chosen as it is the basic component of most routines. Furthermore, it employs the whole subset of instructions used in all the SSE routines. 
\paragraph{}
The table below gives the timing information for all the SSE2 routines. All routines are compared with the corresponding optimized non-sse routine. Each routine was run 1 million times, as the time taken by a single run is too negligible to be recorded properly by a system clock. Additionally, the non-sse code was optimized for performance through techniques such as loop unrolling to avoid overhead. One of the issues with running the same program a million times is that the timing information does not accurately reflect the FSB memory bottleneck mentioned earlier. As all the elements are loaded into the cache in the first run, memory accesses of subsequent runs will be much faster. The compilation was done using gcc with option -O2.

\begin{table}[htdp]
\caption{Execution times of SSE2 routines}
\begin{center}
\begin{tabular}{|| l | p{4cm} | p{4cm} ||}	
	\hline
	{\bf Function}	&	{\bf SSE2 timing (in secs)}	&	{\bf NON-SSE2 timing(in secs)}	\\	\hline
	\verb+add_su3_vector+	&	&	\\	\hline
	\verb+mult_adj_su3_mat_hwvec+	&	8.01	&	10.57	\\	\hline
	\verb}mult_adj_su3_mat_vec}		&	4.03	&	5.68	\\	\hline
	\verb-mult_adj_su3_mat_vec_4dir-	&	15.75	&	25	\\	\hline
	\verb-mult_adj_su3_mat_4vec-	&	15.56	&	25.23	\\	\hline
	\verb-mult_su3_an-	&	12.03	&	17.67	\\	\hline
	\verb-mult_su3_mat_hwvec-	&	8..01		&	10.61	\\	\hline
	\verb-mult_su3_na-	&	12.03	&	20.15	\\	\hline
	\verb-mult_su3_nn-	&	12.03	&	17.69	\\	\hline
	\verb-mult_su3_mat_vec-	&	4.05	&	7.13	\\	\hline
	\verb-mult_su3_mat_vec_sum_4dir-		&	17.49	&	30.01	\\	\hline
	\verb-scalar_mult_add_su3_matrix-	&	3.93	&	7.97	\\	\hline
	\verb-scalar_mult_add_su3_vector-	&			&	\\	\hline
	\verb-su3_projector-	&	4.09	&	16.14	\\	\hline
	\verb-sub_four_su3_vecs-	&	2.33	&	4.44	\\	
	\hline	 
\end{tabular}

\end{center}
\label{default}
\end{table}

For a careful analysis, it would be informative to look at the different kind of SSE2 instructions that have been used. All instructions are broken up into constituent micro-operations by the instruction decoder before they are stored in the trace cache. The complexity and execution time of each instruction depends on the the number of micro-operations that the instruction gets broken down to. The most efficient instructions are those which have less than 6 micro-operations. Basic arithmetic instructions such as addition, multiplication, and logical operations fall in this category, and are the most commonly used instructions in the SSE2 computational routines. 
\paragraph{}
The various move instructions do not present any benefits over the normal non-sse move instructions, as the cost of fetching a 128 bit word is the same. Though the shuffle and unpack operations act on 64-bit words instead of traditional 32-bit words, there is no advantage in terms of efficiency or time to using them in this case. The lack of advantage mainly lies with the fact that the non-sse routines do not have the need for using these instructions. The table below lists the arithmetic instructions and their percentage use in the SSE routines. With MILC, the user has a choice between running in the program in either serial or parallel mode on serial or parallel platforms. In both cases, the improvement in performance is dependent on the percentage of arithmetic SSE instructions in relation to the total number of instructions. The table below gives a break up of the various kinds of instructions that have been used in each routine. 

\begin{table}[htdp]
\caption{Instruction split up of the MILC SSE2 routines}
\begin{center}
\begin{tabular}{|| l | l | l | l | l ||}	
	\hline
	{\bf Function}	&	{\bf Add}	&	{\bf Mul}	&	{\bf Mov}	&	{\bf Shuffle \& others}	\\	\hline
	\verb+add_su3_vector+	&	&	&	&	\\	\hline
	\verb+mult_adj_su3_mat_hwvec+	&	30	&	39	&	53	&	48	\\	\hline
	\verb}mult_adj_su3_mat_vec}		&	15	&	21	&	29	&	24	\\	\hline
	\verb-mult_adj_su3_mat_vec_4dir-	&	65	&	75	&	93	&	96	\\	\hline
	\verb-mult_adj_su3_mat_4vec-	&	63	&	75	&	95	&	96	\\	\hline
	\verb-mult_su3_an-	&	45	&	57	&	77	&	72	\\	\hline
	\verb-mult_su3_mat_hwvec-	&	30	&	39	&	53	&	48	\\	\hline
	\verb-mult_su3_na-	&	45	&	57	&	77	&	72	\\	\hline
	\verb-mult_su3_nn-	&	45	&	57	&	77	&	72	\\	\hline
	\verb-mult_su3_mat_vec-	&	15	&	21	&	29	&	24	\\	\hline
	\verb-mult_su3_mat_vec_sum_4dir-		&	72	&	75	&	113	&	96	\\	\hline
	\verb-scalar_mult_add_su3_matrix-	&	12	&	12	&	33	&	0	\\	\hline
	\verb-scalar_mult_add_su3_vector-	&	&	&	&	\\	\hline
	\verb-su3_projector-	&	9	&	18	&	35	&	36	\\	\hline
	\verb-sub_four_su3_vecs-	&	&	&	&	\\	
	\hline	 
\end{tabular}

\end{center}
\label{default}
\end{table}

\paragraph{}
Tables 2 and 3 give an estimate of the relative efficiency of SSE2 instructions in the different routines used. As the number of SSE2 instructions increases, the percentage improvement also increases, even if there is an increase in the number of mov instructions also. This maybe due to the fact that the prefetching feature of the netburst architecture makes it difficult to quantify the time taken by all the mov instructions. If movs involve loads from adjacent memory locations, the time for the the second mov instruction is much less than the time for the first move, as the second move involves retrieval from cache rather than memory. As SSE2 performs operations in half the time, one would expect that the the speed up with SSE2 would be double that of normal routines. However, this is never observed due to a number of reasons. 
\begin{enumerate}
\item The speed up never matches the ideal case of 2, as all computational instructions are interspersed with data movement to and from memory. 
\item A point to keep in mind is that the nature of the computation requires us to perform some extra instructions such as unpack and  shuffle, which we would not have used in the traditional case. 
\item Since the moves and shifts are interspersed with computational operations, the actual computational throughput is reduced significantly. 
\item Furthermore, the number of registers available for the SSE2 operations is restricted to 8. 
\item There is a further performance hit as computational units have to wait for data to be moved back and forth between cache and SSE2 registers.  
\item One of the problems is that the time for memory retrieval and writes takes a disproportionate amount of time. Hence any improvement in the FSB speed will also increase the overall system as well as SSE performance. 

\end{enumerate}
\paragraph{Serial programs}
For the Serial program case,	\\	 
Time taken = Time for computational routines + time for non computational routines \\ 
Time for computational routines = Time for nonsse instructions + time for sse instructions	\\
Time$_{Normal}$ = Time for normal instructions in comp. routines + time for non-sse instructions in comp. routines + time for non computational routines	\\
Time$_{SSE}$ = Time for SSE instructions in comp. routines + time for non-sse instructions in comp. routines + time for non computational routines
\paragraph{Parallel programs}
For the parallel program case,	\\ 
Time$_{normal}$ = Time for normal instructions in comp. routines + time for non-sse instructions in comp. routines + time for non computational routines + time for communication routines	\\
Time$_{SSE}$ = Time for SSE instructions in comp. routines + time for non-sse instructions in comp. routines + time for non computational routines + time for communication routines	\\
\\
Therefore, in both cases, improvement through SSE routines = Time$_{normal}$ / Time$_{SSE}$	\\
\paragraph{}
For parallel programs the Time$_{Normal}$ and Time$_{SSE}$ have the extra component of communication time. Therefore, the percentage of SSE2 instructions in use goes down. Consequently, in transitioning from serial to parallel programs the SSE2 improvement deteriorates as the non computational times increases. We now have to contend with the communication routine timings. 
\paragraph{}
The highest percentage improvement observed with the use of SSE2, has been with serial code,  specifically 4$^4$ lattices. With parallel jobs, as the lattice size increases, the performance gets better if the number of parallel processes is kept the same. Which means that the job size per node will increase. With an increase in the number of processes, the SSE2 percentage improvement goes down. This is evident from the equation above. 
\paragraph{}
The table below shows the timings for the \verb-ks_imp_dyn1- application of the MILC suite. The program is a serial program and was run on a 2.4 Ghz processor, with an 8 KB L1 data cache, and a 256 KB L2 cache. The system bus speed was 533 Mhz, and RAM size was 2 GB. The compilation was done with gcc, option –O3.

\begin{table}[htdp]
\caption{MILC SSE and SSE2 application times}
\begin{center}
\begin{tabular}{|| l | l | l | l ||}	
	\hline
	{\bf Kind of Application}	&	{\bf Time (in secs)}	&	{\bf Size of lattice}	&	{\bf Type of program}	\\	\hline
	Normal Non-SSE2 (double)	&	220	&	4$^4$	&	Serial	\\	\hline
	SSE2	w/o alignment (double)	&	175	&	4$^4$	&	Serial	\\	\hline
	SSE2	with alignment (double)	&	135	&	4$^4$	&	Serial	\\	\hline	
	Normal Non-SSE (single)	&	153	&	4$^4$	&	Serial	\\	\hline
	SSE	 	w/o alignment (single)	&	115	&	4$^4$	&	Serial	\\	\hline
	SSE		with alignment (single)	&	89	&	4$^4$	&	Serial	\\	\hline	
	Normal Non-SSE (double)	&	17290	&	16$^4$	&	Parallel	\\	\hline
	SSE	 	w/o alignment (double)	&	16614	&	16$^4$	&	Parallel	\\	\hline
	Normal Non-SSE (double)	&	1075		&	8$^4$	&	Parallel	\\	\hline
	SSE	 	w/o alignment (double)	&	1040		&	8$^4$	&	Parallel	\\	\hline
	Normal Non-SSE (single)	&	5367		&	16$^4$	&	Parallel	\\	\hline
	SSE	 	w/o alignment (single)	&	4044		&	16$^4$	&	Parallel	\\	\hline
	\hline	 
\end{tabular}

\end{center}
\label{MILCSSE}
\end{table}

\paragraph{}
One notable fact is that SSE improvements as a percentage of non-SSE improvements are much greater than the corresponding SSE2 improvements. The reason for this becomes clear once we analyze SSE behavior on the light of the serial and parallel program equations above. As SSE operates on 4 operands compared to SSE2 which only operates on 2, the processor is kept busy for a larger percentage of time. Therefore while all the other components of the equation are the same, the time for SSE instructions in computational routines is reduced. Consequently, Time$_{Normal}$/Time$_{SSE}$ is higher.
\section{Further observations}
Programming SSE through assembly instructions is a non-trivial task. To alleviate this difficulty, the Intel compiler offers programmers the flexibility to program the SSE2 instructions through vector classes using c++. Additionally, it is also possible to derive the benefits of using SSE2 or SSE by using the Intel compiler SSE flag option. Our experience showed that explicit assembly level programming showed more improvements than c++ vector class code or compilation with the SSE flag. An important lesson that we learnt from our endeavours is that the cost of implementation has to be weighed against the benefits accruing from it. In this case, do the performance benefits justify amount of time and effort spent in hand coding SSE routines. 

Finally, one of the findings of our study was the amount of dependence that performance has on the memory latency. In our tests, this latency was a direct factor of the processor-FSB speed gap. Several observations led us to the conclusion. With lattice gauge theory simulations, a consistent observation has been the deterioration of speed with larger lattice sizes. This happens due to the fact that the links are flushed frequently from the caches necessitating frequent fetches from memory. For smaller lattice sizes where larger chunks of the lattice fit in memory this phenomenon does not happen this often, as a result of which the timing is much better, as shown in Table~\ref{MILCSSE}. Besides the lowering of efficiency with size, we also observed a lowering of efficiency with reduction of FSB speed. We ran the the MILC algorithms on two identical computers, whose only difference was the speed of the system bus. We observed that the difference in speed exactly matched the difference in Giga floating point operations per second for the two computers. Any improvements which reduce memory latency or the amount of data being fetched have a clear benefit. Lower memory latencies will keep the processor occupied for a greater percentage of time. This improvement in processor efficiency will allow techniques like SSE to achieve their full potential. A promising approach that the MILC algorithms adopt to reducing latencies is to ensure regular memory access patterns. This approach exploits the prefetching feature that most modern processors share. As subsequent memory address contents are going to be transferred to cache ahead of their accesses, the memory latency is reduced, implying faster processing times.
\paragraph{}
Our experience also showed us the advantages of using aligned data and consequently aligned SSE2 instructions. The improvement with alignment was observed in SSE as well as non sse cases, and should be an important factor for consideration. The benefits of using alignment are shown more clearly in Table~\ref{align}.

\begin{table}[htdp]
\label{align}
\caption{Effect of alignment}
\begin{center}
\begin{tabular}{|| l | l | p{3cm} | p{3cm} ||}	
	\hline
	{\bf Kind of Application}	&	{\bf Size of lattice}	&	{\bf Time with alignment (in secs)}	&	{\bf Time w/o alignment (in secs)}	\\	\hline
	SSE2 parallel	&	8 x 4$^3$	&	23.9		&	31.1	\\	\hline
	SSE2 parallel	&	16 x 8 x 4$^2$	&	39.73	&	45.18	\\	\hline
	SSE2 parallel	&	32$^2$ x 4$^2$	&	314		&	323	\\	\hline
	SSE2 parallel	&	8$^2$ x 4$^2$	&	36		&	40.3	\\	\hline	 
\end{tabular}
\end{center}
\end{table}

\section{Conclusions}
\paragraph{}
Our motivation in undertaking the study was to improve the performance of our lattice gauge theory simulations. From Tables~\ref{MILCSSE} and~\ref{align}, SSE improves performance. The improvements, however, are less than the maximum gain possible. Future projects which consider performance improvements through the SSE or for that matter any toolkit or framework should consider the factors which play a role in performance. For SSE, the following factors were important:
\begin{enumerate}
\item What percentage of execution time is taken up by computational routines. For lattice gauge theory simulations, the time taken by the computational routines monopolizes execution time.
\item Is the percentage of SSE instructions in considering all computational instructions, high? The number of SSE instructions as a percentage of the total number of computational instructions is large.
\item Architectural constraints such as the FSB bottleneck.
\end{enumerate}
\paragraph{}
Ultimately, the power of SSE2 should be weighed against factors such as cost of development and the percentage of likely improvement. As we have shown, SSE is best for applications that share characteristics such as 
\begin{enumerate}
\item Large percentage of computations
\item Regular memory access patterns
\item Architectures which have high bus speeds that reduce the imbalance between processor speed and bus speed
\end{enumerate}
\paragraph{}
The impact of SSE should be weighed in with all these factors, and also against possible system enhancements in the future. It is important to ask if a particular computational technique justifies the time and effort needed. The equations from Section 5 were used for judging the effect of the factors mentioned above. We believe that the improvement with SSE was significant enough to merit the effort. We believe that the model for SSE may apply equally to other computational techniques for speeding up simulation performance time. Systems may vary and so will implementations but the methodology used for analysis is platform independent and could be extended to applications on different platforms.

\end{document}